\documentclass[twocolumn,prl]{revtex4}
\usepackage{graphicx,bm}
\newcommand{\be}{\begin{equation}}
\newcommand{\ee}{\end{equation}}
\newcommand{\bea}{\begin{eqnarray}}
\newcommand{\eea}{\end{eqnarray}}

\renewcommand{\phi}{\varphi}
\renewcommand{\epsilon}{\varepsilon}

\begin{document}

\title{Nonlocal Charge Transport Mediated by Spin Diffusion in the Spin-Hall 
Effect Regime}
\author{D. A. Abanin,${}^1$ A. V. Shytov,${}^2$ L. S. Levitov,${}^1$ B. I. Halperin${}^3$}
\affiliation{
${}^1$ 
 Department of Physics,
 Massachusetts Institute of Technology, 77 Massachusetts Ave,
 Cambridge, MA 02139\\
${}^2$ 
Brookhaven National Laboratory, Upton, 
New York 11973-5000\\
${}^3$
Lyman Laboratory, Physics Department, Harvard University, 
Cambridge MA 02138
}

\begin{abstract}

A nonlocal electric response in the spin-Hall regime,
resulting from spin diffusion mediating charge conduction,
is predicted.
The spin-mediated transport stands out due to its long-range character,
and can give dominant contribution to nonlocal resistance.
The characteristic range of nonlocality, 
set by the spin diffusion length, can be large enough to 
allow detection of this effect in materials such as GaAs
despite its small magnitude. 
The detection is facilitated by a characteristic nonmonotonic dependence 
of transresistance
on the external magnetic field, exhibiting sign changes and decay.
\end{abstract}

\maketitle

The spin Hall effect (SHE) is a phenomenon arising due to spin-orbit coupling 
in which charge current passing 
through a sample leads to spin transport in the transverse direction 
\cite{Dyakonov71}. This phenomenon has been attracting continuous interest, 
partially because of the rich 
physics and diversity of SHE mechanisms\,\cite{Engel2006} 
and partially because 
SHE enables generating spin polarization on a micron scale and electrical detection of spin-polarized currents, which are the key ingredients of 
spintronics \cite{Wolf01}. 
Theoretically, two main types of SHE have been studied, (i) extrinsic, 
being due to spin-dependent scattering on impurities \cite{Dyakonov71} and 
(ii) intrinsic, arising from the spin-orbit terms in the band Hamiltonian \cite{Sinova04}.  
Both extrinsic and intrinsic SHE have been detected experimentally 
\cite{Kato04, Wunderlich05, Sih06, Stern06} using optical techniques. 
Reciprocal SHE (that is, transverse voltage induced by a spin-polarized 
current) was observed in Al nanowire \cite{Valenzuela06}, 
where ferromagnetic contacts were used to inject spin-polarized current into 
the sample, and in Pt film \cite{Saitoh2006}.  

Here we show that the SHE relation between charge current and spin current
leads to an interesting spin-mediated nonlocal charge transport,
in which spins generated by SHE diffuse 
through the sample and, by reverse SHE, induce electric current
elsewhere. The range of nonlocality of this charge transport
mechanism
is of the order of 
the spin diffusion length $\ell _s=\sqrt{D_s\tau_s}$, 
where $D_s$ is the spin diffusion constant, 
and $\tau_s$ is the spin relaxation time.
%
The observation of such nonlocal charge transport due to SHE can be made fully electrically, 
which represents a distinct advantage compared to the methods 
relying on the sources of 
spin-polarized current\,\cite{Valenzuela06,Saitoh2006}. Although the nonlocal electrical signal, 
estimated below, is small, by optimizing the multiterminal geometry one can
enhance the nonlocal character of the effect and easily
distinguish it from the ohmic transport.

The main distinction of the spin-mediated electrical effect
considered in the present work
from related ideas  
discussed earlier\,\cite{Levitov85,Hirsch99,Hankiewicz2004,Dyakonov2007},
is that here we identify a situation 
in which the spin-mediated charge transport, 
due to its nonlocality, {\it dominates} over
the ohmic contribution.
In particular, Refs.\cite{Levitov85,Dyakonov2007} considered a correction to
the bulk conductivity resulting from spin diffusion and SHE,
which was small in magnitude and thus could only be made detectable by
its characteristic contribution to magnetoresistance.
In Ref.\cite{Hankiewicz2004} 
a spin-assisted electrical response was predicted in
a multiterminal mesoscopic system, studied numerically
in a weak disorder regime with the mean free path
comparable to the system size.
In this case the spin-dependent fraction of the multiterminal 
Landauer-B\"uttiker conductance is also small.

\begin{figure}
\includegraphics[width=3.1in]{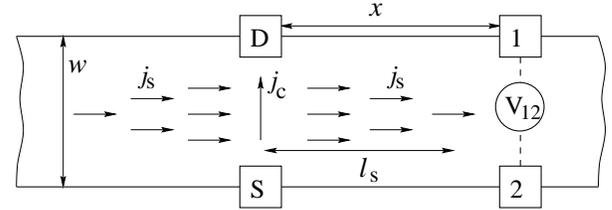}
\vspace{-3mm}
\caption[]{
Nonlocal spin-mediated charge transport schematic. 
Charge current $j_c$ applied across 
a narrow strip generates, via SHE, a longitudinal spin current $j_s$.
After diffusing over 
a distance $x\sim\ell_s \gg w$, the spins induce 
a transverse voltage $V_{12}$ on the probes 1 and 2
via the 
reciprocal SHE. 
For a narrow strip, $w\ll\ell_s$, 
the spin-mediated nonlocal contribution 
exceeds the ohmic contribution
that decays as $e^{-\pi|x|/w}$.
}
\label{fig0}
\end{figure}

The geometry that will be of interest to us  
is a strip of width $w$ narrow compared to the spin diffusion length $\ell_s$,
with current and voltage probes attached as illustrated in Fig.\ref{fig0}. 
The nonlocal charge transport manifests itself in a voltage 
$V_{12}$ across the strip measured 
at distances $x\sim \ell_s$ from the current source (S) and drain (D).
Since in the purely ohmic, non-SHE regime
the voltage away from the source decays 
exponentially on the length scale $w/\pi$, 
our nonlocal spin-dependent voltage can be
easily made to exceed the ohmic spin-independent contribution.
Furthermore, the spin-mediated effect will exhibit a characteristic
oscillatory dependence on the in-plane magnetic field 
arising due to spin precession during transport.
At a distance $|x|\sim \ell_s$ from the source it
will oscillate and 
decay as a function of magnetic field on the scale $\omega_B\sim 1/\tau_s$, 
where $\omega_B$ is the electron spin Larmor frequency.


We consider, as a simplest case, an infinite narrow strip 
\[ 
-\infty<x<+\infty,\quad -w/2<y<w/2, \quad
(w\ll\ell_s),
\]
as illustrated in Fig.\ref{fig0}.
We assume, without loss of generality,
that current source and drain leads are narrow, 
connected to the sample at the points
$(0,\pm {\textstyle\frac12}w)$.

The electric potential in such a sample, described by ohmic conductivity
$\sigma$, can be 
found as a solution of the 2D Laplace's 
equation with the boundary condition
$
j_{y}(x,y=\pm {\textstyle\frac12}w)=I\delta(x)
$,
where $I$ is the external current. 
Solving it by the Fourier method, 
we find
%
\be\label{eq:potential_dist}
\phi(x,y)=-\int dk\frac{I e^{ikx}\sinh (ky)}{2\pi\sigma k \cosh({\textstyle\frac12}kw)}
.
\ee
%
%
In what follows we will need the electric field tangential component 
at the boundaries of the strip, 
$E_{x,\pm}(x)=-\partial _x \phi_{\pm}(x)$, where 
$+$ and $-$ signs correspond to the strip upper edge ($y=+w/2$) 
and lower edge ($y=-w/2$). 
This component of the electric field is
found from the potential (\ref{eq:potential_dist}) at the 
boundaries of the sample as
%
%
%
\be\label{eq:E_x3}
E_{x,\pm}(x)=\mp \frac{2I {\rm sign}x}{w\sigma} \sum _{{\rm odd}\, n>0} 
e^{-n|\tilde x|}
= \mp  \frac{I}{w\sigma\sinh \tilde x}
,
\ee
where $\tilde x=\pi x/w$.
Potential difference between the strip edges, $\Delta\phi=\phi_+(x)-\phi_-(x)$,
evaluated in a similar way, decreases as $e^{-\pi|x|/w}$ at $|x|\gtrsim w$.

Below we focus on the case of extrinsic SHE, when 
the $k$-linear Dresselhaus and Rashba terms in electron spectrum 
are negligible. Then the
spin fluctuation generated by SHE evolves according to
the diffusion equation, 
\be\label{eq:spin_diff}
D_s \partial ^2 s(x,y)-\Xi(x,y)-\frac{1}{\tau_s}s(x,y)=0,
\ee
where $s$ is the $z$-component of the spin density, 
$D_s$ is the spin diffusion coefficient, $\tau_s$ is the spin relaxation time. The source term $\Xi$ in Eq.(\ref{eq:spin_diff}) 
describes spin current arising due to the spin Hall effect:
\be\label{eq:gamma}
\Xi_\gamma(x,y)=\nabla. j_s=\partial_{\alpha} (\beta_{s} \epsilon_{\alpha \beta\gamma} E_\beta(x,y)),
\ee 
where $\beta_s$ is the spin Hall conductivity. 
In the presence of the $k$-linear Dresselhaus and/or Rashba interaction,
spin transport is more complicated due to SO-induced
precession and dephasing\,\cite{Mishchenko04,Tse06-2}.
The modification of the nonlocal electric effect in this case
will be briefly discussed at the end of the paper.

Since $\nabla\times {\bf E}=0$, the 
spin current source (\ref{eq:gamma}) vanishes in the bulk and is only non-zero 
at the strip boundaries: 
\be\label{eq:bound_source}
\Xi (x,y)=  \beta _s \delta(y-{\textstyle\frac12}w) E_{x,+}(x)-\beta _s \delta(y+{\textstyle\frac12}w) E_{x,-}(x). 
\ee
%
The distinction from the ohmic contribution becomes most clear when
our strip is relatively narrow,
$w\ll \ell_s$.
%
In this case the spin current and spin density induced by SHE
are approximately constant across the strip.
Thus we can integrate over $y$ and solve a one-dimensional 
spin diffusion problem. 
Suppressing the $y$ dependence in Eq.(\ref{eq:spin_diff}), 
we take $\Xi(x)=\beta_s E_{x,+}(x)-\beta_s E_{x,-}(x)$. 

Solution of Eq.(\ref{eq:spin_diff})
in the Fourier representation reads: 
\be\label{eq:s_density}
s_k=-
\frac{p_k
}{D_s k^2+1/\tau_s}
,\quad
p_k= \frac{1}{2\pi}\int_{-\infty}^{+\infty} dx \Xi (x) e^{-ikx}
. 
\ee
This expression can be simplified by noting that $\Xi(x)$ is an odd function of $x$ and that the integral over $x$
converges at $|x|\lesssim w$, while we are interested in 
the harmonics with much lower $k\sim 1/\ell_s \ll 1/w$.
Sending $k$ to zero in the integral, we obtain
 \be\label{eq:s_density3}
s_k=\frac{1}{\pi}\frac{ik G}{D_s k^2+1/\tau_s},  
\ee
where the spin dipole $G$ is given by
\be\label{eq:G2}
G=\int_0^\infty  \Xi(x) x\, dx  = -\frac{I\beta_s w}{2\sigma}
\ee
(we used Eqs.(\ref{eq:E_x3}),(\ref{eq:bound_source}) to evaluate
this expression).

Now we can find the spin current 
\be\label{eq:spin_current}
J_s(x)=-D_s\partial_x s(x)
,
\ee
where the spin density $s(x)$ is obtained 
by the inverse Fourier transform of 
Eq.(\ref{eq:s_density3}). 
Using Eq.(\ref{eq:G2}),
we find 
\be\label{eq:spin_current2}
J_s(x)=\frac{I\beta_s w}{2\sigma \ell_s} e^{-|x|/\ell_s}
. 
\ee
(this expression is valid for $x$ not too close to the
source, $|x|\gtrsim w$).
The expression (\ref{eq:spin_current2}) gives the net spin current 
rather than the spin current density, 
as we have been solving a 1D diffusion problem. 
This spin current generates a voltage across the sample
\be\label{eq:nonlocal_voltage}
\delta V(x)=\frac{\beta_c J_s(x)}{\sigma}=\frac{I\beta_c \beta_s w}{2\ell_s \sigma^2} 
e^{-|x|/\ell_s},
\ee
where $\beta_c$
describes charge current arising in response to spin current, 
$j_c^\alpha=\beta_c \epsilon_{\alpha\beta}j_s^\beta$. 
Relating $\beta_c$ to the spin Hall conductivity as $\beta_c=\beta_s/\sigma$,
we write the nonlocal response (\ref{eq:nonlocal_voltage})
as a transresistance
%
\be\label{eq:nonlocal_voltage2}
R_{\rm nl}(x)=\frac{\delta V(x)}{I}
=\frac{1}{2} \left(\frac{\beta_s}{\sigma}\right)^2 
 \frac{w}{\sigma l_s}
e^{-|x|/\ell_s}.
\ee
We emphasize that for the extrinsic SHE\,\cite{Dyakonov71}, 
the spin current is established on the length scale of the order of the electron mean free path $\ell$, here taken to be much smaller than 
the strip width $w$. Thus the inhomogeneity of the charge current $j_c$ 
(Fig.\ref{fig0}) on the length scale set by $w$
does not affect our analysis.  
The estimate (\ref{eq:nonlocal_voltage2}) for the nonlocal voltage 
is therefore accurate as long as $w\gg \ell$.

We now compare the magnitude of the nonlocal contribution 
(\ref{eq:nonlocal_voltage2}) for several materials 
where extrinsic SHE has been observed.
For the transresistance (\ref{eq:nonlocal_voltage2}) to be large, 
one would like to have a material with a large ratio $\beta_s/\sigma$, 
and a large spin diffusion length $\ell_s$. 
For Si-doped GaAs with electron density 
$n=3\times 10^{16}\, {\rm cm}^{-3}$, the 3D charge conductivity,
spin Hall conductivity, and spin diffusion length, reported in 
\cite{Sih06,Kato04}, are given by 
$\sigma_{3D}\approx 2.5\times 10^{-3}\, \Omega^{-1} \mu {\rm m}^{-1}, \,\,
{\beta_s}_{3D} \approx 5\times 10^{-7} \, \Omega^{-1}\mu{\rm m}^{-1}, \,\,
\ell _s \approx 9\, \mu{\rm m}$. 
Our two-dimensional quantities $\sigma$, $\beta_s$ are related to 
the 3D quantities as 
$\sigma=\sigma_{3D}w_z$, $\beta_s={\beta_s}_{3D} w_z$, 
where $w_z$ is the sample thickness.
Taking $w_z=2\, \mu{\rm m}$\,\cite{Sih06,Kato04},  
and choosing the sample width to be $w=0.5\, \mu{\rm m}$, 
we estimate the 
transresistance (\ref{eq:nonlocal_voltage2}) as
\be\label{eq:R_nonlocal}
R_{\rm nl}(x)
\approx 2\times 10^{-7} \times e^{-|x|/\ell_s}\, [{\rm Ohm}]. 
\ee
Although small, it
by far exceeds the ohmic conduction contribution which
at a distance $x$ is proportional to $\sigma^{-1}e^{-\pi |x|/w}$.
Indeed, for a typical $x\approx\ell_s$ the ohmic contribution
is negligibly small: 
$e^{-\pi\ell_s/w}\approx e^{-57}\approx 10^{-24.8}$. 

In the case of InGaAs, the 3D charge conductivity and 3D spin Hall conductivity have values similar to those quoted above for GaAs (see Ref.\,\cite{Kato04}), 
$\sigma_{3D}\approx 2.5\times 10^{-3}\, \Omega^{-1} \mu {\rm m}^{-1}, \,\,
{\beta_s}_{3D} \approx 5\times 10^{-7} \, \Omega^{-1}\mu{\rm m}^{-1},$ while spin diffusion length is considerably shorter, 
$\ell _s \approx 2\, \mu{\rm m}$. Therefore, in order for the nonlocal voltage 
(\ref{eq:nonlocal_voltage2}) at $|x|\sim \ell_s$ 
to exceed the ohmic contribution, proportional to $\sigma^{-1}e^{-\pi|x|/w}$, 
the sample width $w$ must satisfy $w\ll \pi \ell_s/(\, 2{\rm ln}(\sigma_{3D}/\beta_{3D}))\approx 360 \, {\rm nm}$. 

Another material exhibiting extrinsic SHE is ZnSe \cite{Stern06}. 
For carrier concentration $n=9\times 10^{18}\,{\rm cm^{-3}}$ the 3D charge and spin Hall conductivities are given by $\sigma_{3D}\approx 2\times 10^{-1}
\Omega^{-1}\mu {\rm m}^{-1}$, 
${\beta_s}_{3D} \approx 3\times 10^{-6} \Omega^{-1}\mu{\rm m}^{-1}$, 
having the ratio ${\beta_s}_{3D}/\sigma_{3D}\approx 1.5\times 10^{-5}$ 
about ten times smaller than in GaAs and InAs. 
The spin diffusion length in this material is comparable to that in InAs, $\ell_s\approx 2\,{\rm \mu m}$. 

Extrinsic SHE has been also demonstrated in metals, Al (see Ref.\,\cite{Valenzuela06}) and Pt (see Ref.\,\cite{Kimura2007,Guo2007}). In Al, ${\beta_s}_{3D}\approx 3\times 10^{-3}\Omega^{-1} \mu {\rm m}^{-1}$, the ratio of spin Hall and charge conductivities is 
${\beta_s}_{3D}/\sigma_{3D}\approx 1\times 10^{-4}$, while the spin diffusion length is $\ell_s\approx 0.5\,{\rm \mu m}$. Therefore, to separate 
the spin effect
from the ohmic contribution, one needs to fabricate samples with $w\ll \pi\ell_s/(2\, {\rm ln}(\sigma_{3D}/\beta_{3D}))\approx 85 \, {\rm nm}$. 
Although the ratio ${\beta_s}_{3D}/\sigma_{3D}\approx 0.37$ is large in Pt, 
the observation of the nonlocal effect in this material 
is hindered by its extremely small spin diffusion lengh, 
$\ell_s \approx 10\,{\rm nm}$ \cite{Kurt02}. 
We therefore conclude that GaAs systems seem to provide an optimal combination
of parameter values for the observation of the nonlocal transport.

We now analyze the effect of an in-plane magnetic field on 
the transresistance (\ref{eq:R_nonlocal}).
In the presence of magnetic field, the spin diffusion equation (\ref{eq:spin_diff}) is modified as  
\be\label{eq:spin_diffB}
D_s \partial ^2 {\bf s}- {\bf \Xi}-\frac{{\bf s}}{\tau_s}+ 
\left[{\bf  \omega}_B \times {\bf s} \right] =0,
\ee
where ${\bf \omega}_B=g\mu_B {\bf B}$ is the Larmor precession frequency, 
$\mu_{\rm B}=9.27\times 10^{-24}\, 
{\rm J/T}$ is the Bohr magneton and $g$ is the $g$-factor. 
As we shall see below, the interesting field range is 
$\omega_B\lesssim D_s/w^2$. Since in this case the variation
of spin polarization across the strip is negligible,
we can again integrate over 
the $y$ coordinate and solve a one-dimensional diffusion problem. 

For the magnetic field parallel to the $x$ axis, 
Eq.(\ref{eq:spin_diffB}) takes the following form in the 
Fourier representation:
\be\label{eq:spin_diff_momentum}
 \left[\begin{array}{ccc}
         g(k) & 0 & 0 \\
          0 &  g(k)   &  \omega_B             \\
          0     &  -\omega_B   &    g(k)  
      \end{array} 
 \right] 
\left(  \begin{array}{c}
       s_k^x \\ 
       s_k^y \\
       s_k^z
        \end{array}
\right)=- 
\left(  \begin{array}{c}
       \Xi_k^x \\ 
       \Xi_k^y \\
       \Xi_k^z
        \end{array}
\right), 
\ee
where $ g(k)=D_s k^2+1/\tau_s $.
For the situation of interest, when only the $z$ component of 
the source ${\bf \Xi}$ is nonzero, the solution for 
$s^z$ 
is given by 
\be\label{eq:s_k_magnetic}
s_k^z=-\frac{\Xi_k^z(D_s k^2+1/\tau_s)}{(D_sk^2+1/\tau_s)^2+\omega_B^2}. 
\ee  
Following the same steps as in the absence of magnetic field 
(notice that the source term $\Xi ^z$ is not affected by the 
magnetic field), we 
obtain the spin current, 
\be\label{eq:spin_currentB}
J_s(x)=\frac{I\beta_s w}{2\sigma} {\rm Re}\left[ q_+e^{-q_+ |x|} \right], 
\ee
where $q_+={\sqrt{1+i\omega_B \tau_s}}/{\ell _s}$. 

\begin{figure}
\includegraphics[width=3.4in]{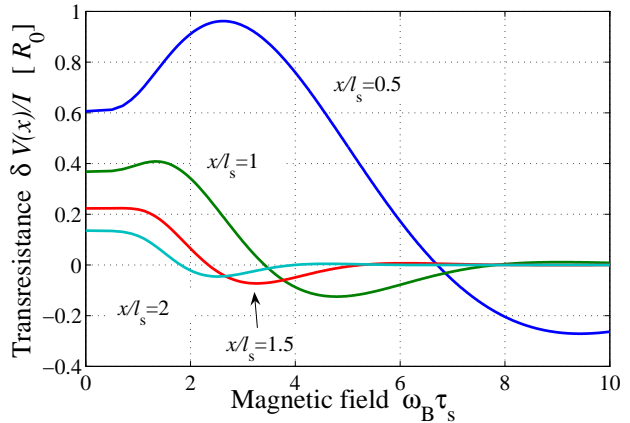}
\vspace{-7mm}
\caption[]{Nonlocal resistance $R_{\rm nl}=\delta V(x)/I_{sd}$,
Eq.(\ref{eq:nonlocalB}), 
in units of $R_0=\beta_c\beta_sw/2\sigma^2\ell_s$
{\it vs.} the in-plane magnetic field for several values of $x$
(see Fig.\ref{fig0}).
[Here $\omega _B$ and $\tau_s$ are the electron spin
Larmor precession frequency $\omega _B$ 
and dephasing time.]
The nonlocal response {\it increases} at weak
fields, $\omega_B\tau_s \lesssim \ell _s/|x|$, 
changes sign at $\omega_B\tau_s \sim \ell _s/|x|$,
and is suppressed at 
$\omega_B\tau_s\gtrsim \ell _s/|x|$, simultaneously exhibiting oscillations. }
\label{fig2}
\end{figure}

The nonlocal response due to the voltage induced by the 
spin current (\ref{eq:spin_currentB}),
found as $\delta V (x)=\frac{\beta_c}{2\sigma} J_s(x)$,
is 
\be\label{eq:nonlocalB}
R_{\rm nl}(x)=\frac{\delta V (x)}{I}
=\frac{\beta_c \beta_s w}{2\sigma^2}
{\rm Re}\left[ q_+e^{-q_+ |x|} \right]. 
\ee
This expression is simplified in the limit of strong magnetic field, 
$\omega_B \tau_s\gg 1$, by factoring an oscillatory 
term:
\be\label{eq:nonlocal_strongB}
R_{\rm nl}(x)=\frac{  I\beta_c \beta_s w \eta}
{\sqrt 2\ell_s \sigma^2} {\rm sin}\left(\frac{\eta |x|}{ \ell_s}+\frac{\pi}{4}    \right) e^{-\eta |x|/{ \ell_s} }, 
\ee
where $\eta=\sqrt{\omega_B \tau_s/2}$. 
Compared to 
the result found in the absence of magnetic field, 
Eq.(\ref{eq:nonlocal_voltage}), 
the nonlocal voltage $\delta V(x)$ 
is amplified by a factor of $\sqrt 2\, \eta$,
decaying 
on a somewhat shorter length scale 
$\tilde{\ell}=\ell_s/\eta$.

The dependence of $R_{\rm nl}$, Eq.(\ref{eq:nonlocalB}), on the 
in-plane magnetic field 
is illustrated in Fig.\ref{fig2}. 
Enhancement of $\delta V$ at weak fields,
$\omega_B\tau_s\lesssim 1$, is followed by a sign change
at $\omega_B\tau_s\sim 1$, and suppression at $\omega_B\tau_s\gtrsim 1$. 
The zeros of $\delta V$
can be found approximately for $\ell_s\gtrsim |x|$ from Eq.(\ref{eq:nonlocal_strongB}):
\[
\omega_{B,n} \tau _s \approx 2\pi^2(n-1/4)^2\ell_s/|x| , 
\]
with integer $n>0$. (The condition $\ell_s/|x|\gtrsim 1$ ensures that 
$\omega_n \tau_s\gg 1$, necessary for Eq.(\ref{eq:nonlocal_strongB}) to be valid.)


For GaAs, $g=-0.44$ and $\tau_s\sim 10\, {\rm ns}$  (see Ref.\cite{Sih06}), 
and therefore the field necessary to observe 
the oscillations and suppression of $R_{\rm nl}$ at $|x|\sim \ell_s$, is quite weak:
%
\be\label{eq:mag_field}
B\sim B_* = \frac{\hbar}{ g \mu_{\rm B}\tau _s}\approx 2\, {\rm mT}. 
\ee
The transresistance measured at $|x|=\ell_s$ will change sign at the fields 
$B\approx 11.1\, {\rm mT}, \, 60.4\, {\rm mT},...$,
decreasing in magnitude
as illustrated in Fig.\ref{fig2}. 

Nonlocal electric transport can result not only 
from the extrinsic spin scattering
mechanisms discussed above but also from the intrinsic spin-orbital effects.
Below we briefly consider this effect and present a rough 
estimate for a 2D electron gas with 
Rashba spin-orbit coupling, $H_{SO} = \alpha {\hat z}.[{\bm \sigma} \times {\bf p}]$, 
where~${\bm \sigma}$ and  ${\bf p}$ are electron spin and momentum, 
$\hat z$ is the unit normal vector,
and~$\alpha$ is SO interaction constant 
({\it cf.} Ref.\cite{Hankiewicz2004}.)
Potential scattering by impurities leads to Dyakonov-Perel spin relaxation
with spin diffusion length~$\ell_s = \hbar / m_\ast \alpha$,  
where~$m_\ast$ is the effective electron mass.
In such a system, unlike extrinsic SHE, electric
field induces spin density
rather than spin current~\cite{Levitov85,Edelstein90}. 
The in-plane spin polarization  ${\bf s}\propto {\hat z}\times {\bf j}_c$, 
induced by the source-drain current,
will diffuse along the strip (Fig.\ref{fig0}),
forming a profile $s(x) \sim \alpha m_\ast \tau e^{-|x|/\ell_s} I / \sigma$.
(For an estimate we used the spin diffusion equation derived
in~\cite{Mishchenko04}, Eq.(13), 
which has a form similar to Eq.~(\ref{eq:spin_diffB}) 
with a source 
term ${\bf \Xi}(x,y) \sim \alpha^3 (p_F \tau)^2 e m_\ast\sigma^{-1} {\hat z}\times{\bf j}_c(x,y)$, 
where $p_F$ is Fermi momentum, and $\tau$ is the momentum relaxation time.)

By Onsager relation, the spin density  
creates electric current, ${\bf j}' \sim e \alpha^3 p_F^2 \tau {\hat z}\times {\bf s}$, giving rise to
transresistance
\begin{equation}
R_{\rm nl} (x)
\sim 
\left(\frac{\hbar \tau}{m_\ast \ell_s^2}\right)^2 
\frac{w}{\sigma \ell_s} e^{-|x|/\ell_s}
. 
\end{equation}
For a GaAs quantum well with electron density $n = 10^{12} {\rm cm}^{-2}$,
mobility $\mu = 10^5 {\rm cm}^2/Vs$, 
spin orbit splitting $\alpha p_F = 100\mu{\rm V}$, 
and width $w = 0.5\mu {\rm m}$, we estimate $R_{\rm nl} \sim 10^{-5}$Ohm,  
which is somewhat larger than the ohmic contribution $\sim 10^{-6}$Ohm 
at $x = \ell_s \approx 3\mu{\rm m}$. 
However, larger values of the mean free path
in quantum wells $\sim 1\mu{\rm m}$ enhance nonlocality of the ohmic
contribution. This may hinder observation of spin-mediated transport
despite a larger value of $R_{\rm nl}$ 
({\it cf.} Ref.\cite{Hankiewicz2004}). 


In conclusion, spin diffusion in the SHE regime can give rise
to nonlocal charge conductivity. A relatively large nonlocality scale,
set by the spin diffusion length, can be used to separate 
the spin-mediated transresistance from the ohmic conduction effect.
In a narrow strip geometry,
the transresistance has a nonmonotonic dependence on 
the external in-plane magnetic field, exhibiting multiple 
sign changes and damping.
Our estimates indicate that observation of the nonlocal conductivity 
is possible for currently available $n$-doped GaAs samples. 
 
We benefited from 
discussions with M. I. Dyakonov and  H.-A. Engel. 
This work is supported by NSF MRSEC (grant DMR 02132802), 
NSF grants DMR 0541988 and PHY 0646094, 
and DOE (contract DEAC 02-98 CH 10886).


\end{document}